\documentclass[sigconf]{acmart}

\usepackage{bm}

\newcommand{\x}{{\bf x}}

\newcommand{\z}{{\bf z}}

\AtBeginDocument{%
  \providecommand\BibTeX{{%
    \normalfont B\kern-0.5em{\scshape i\kern-0.25em b}\kern-0.8em\TeX}}}

\setcopyright{acmcopyright}
\copyrightyear{2021}
\acmYear{2021}
\setcopyright{acmcopyright}\acmConference[CIKM '21]{Proceedings of the 30th ACM International Conference on Information and Knowledge Management}{November 1--5, 2021}{Virtual Event, QLD, Australia}
\acmBooktitle{Proceedings of the 30th ACM International Conference on Information and Knowledge Management (CIKM '21), November 1--5, 2021, Virtual Event, QLD, Australia}
\acmPrice{15.00}
\acmDOI{10.1145/3459637.3481941}
\acmISBN{978-1-4503-8446-9/21/11}

\begin{document}
\fancyhead{}
\title{One Model to Serve All: Star Topology  Adaptive Recommender for Multi-Domain CTR Prediction}





\author{Xiang-Rong Sheng, Liqin Zhao}
\authornote{Xiang-Rong Sheng and Liqin Zhao contributed equally to this work. 
\\ \{xiangrong.sxr, liqin.zlq\}@alibaba-inc.com}
\author{Guorui Zhou, Xinyao Ding, Binding Dai, Qiang Luo \\Siran Yang, Jingshan Lv, Chi Zhang,  Hongbo Deng, Xiaoqiang Zhu}
\affiliation{%
  \institution{Alibaba Group}
  \city{Beijing, China}
}
\renewcommand{\shortauthors}{Sheng and Zhao, et al.}

\begin{abstract}
Traditional industry recommendation systems usually use data in a single domain to train models and then serve the domain. However, a large-scale commercial platform often contains multiple domains, and its recommendation system often needs to make click-through rate (CTR) predictions for multiple domains. Generally, different domains may share some common user groups and items, and each domain may have its own unique user groups and items. Moreover, even the same user may have different behaviors in different domains. In order to leverage all the data from different domains, a single model can be trained to serve all domains. However, it is difficult for a single model to capture the characteristics of various domains and serve all domains well. On the other hand, training an individual model for each domain separately does not fully use the data from all domains. In this paper, we propose the Star Topology Adaptive Recommender (STAR) model to train a single model to serve all domains by leveraging data from all domains simultaneously, capturing the characteristics of each domain, and modeling the commonalities between different domains. Essentially, the network of each domain consists of two factorized networks: one centered network shared by all domains and the domain-specific network tailored for each domain. For each domain, we combine these two factorized networks and generate a unified network by element-wise multiplying the weights of the shared network and those of the domain-specific network, although these two factorized networks can be combined using other functions, which is open for further research. Most importantly, STAR can learn the shared network from all the data and adapt domain-specific parameters according to the characteristics of each domain. The experimental results from production data validate the superiority of the proposed STAR model. Since late 2020, STAR has been deployed in the display advertising system of Alibaba, obtaining 8.0\% improvement on CTR and 6.0\% increase on RPM (Revenue Per Mille).
\end{abstract}
\ccsdesc[500]{Information systems~Information retrieval}
\keywords{Multi-Domain Learning, Recommender System, Display Advertising}




\maketitle
\begin{figure}[!t]
    \centering
    \includegraphics[width=\columnwidth]{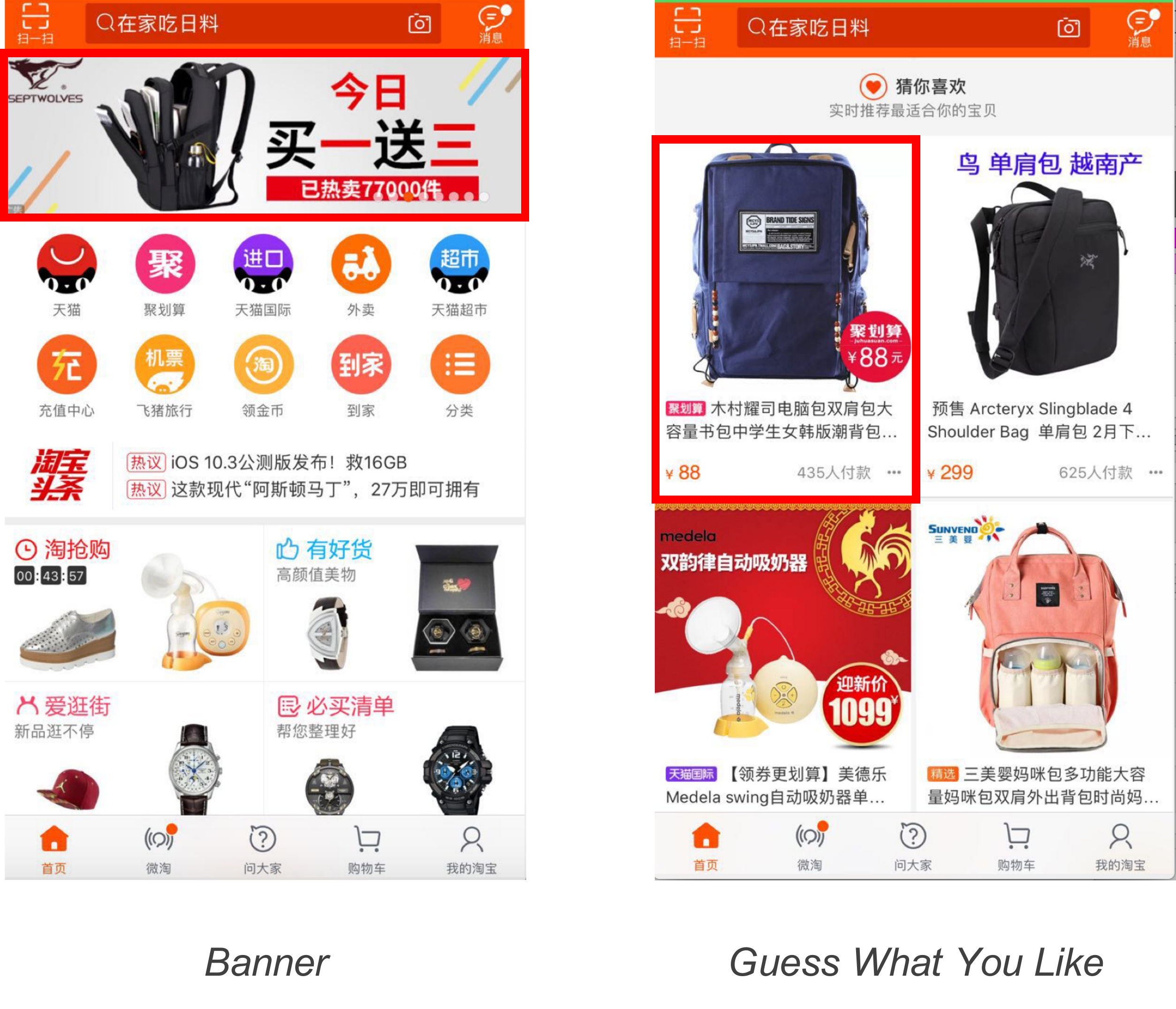}
    \caption{Two representative business domains, \textit{Banner} and \textit{Guess What You Like} on Taobao mobile app home. A business domain is referred to as a specific spot that items are presented to users in the mobile app and PC websites.}
    \label{fig:business_domain}
\end{figure}
\section{Introduction}\label{sec:intro}
Traditional CTR prediction models~\cite{zhou2018din,zhou2019dien,rendle2010factorization,guo2017deepfm,cheng2016wide} focus on single-domain prediction, where the CTR model serves for a single business domain after trained with examples collected from this domain. Each business domain is a specific spot that items are presented to users on the mobile app or PC websites. At large commercial companies like Alibaba and Amazon, there are often many business domains that need CTR prediction to enhance user satisfaction and improve business revenue. For example, in Alibaba, the business domains range from \textit{Guess What You Like} in Taobao App homepage, \textit{Banner} of Taobao App homepage to other domains~\cite{ZhuJTPZLG2017OCPC}. Figure~\ref{fig:business_domain} shows two representative business domains in Alibaba.
\begin{itemize}
    \item \textit{Banner}: In \textit{banner}, the items to be recommended appears in the top banner of the Taobao home page. The item can be a single commodity, a store, or a brand.
    \item \textit{Guess What You Like}: In \textit{Guess What You Like}, items are all single commodities and displayed to users in the left or right column.
\end{itemize}
Since different business domains have overlapping user groups and items, there exist commonalities among these domains. Enabling information sharing is beneficial for learning the CTR model of each domain.
However, the specific user group may be different and the users' behaviors also change in various domains. These distinctions result in domain-specific data distributions. Simply mixing all the data and training a single shared CTR model can not work well on all domains.

Besides mixing data and training a shared model,  another simple solution is to build a separate model per business domain. This strategy also has some downsides: (1) some business domains have much less data than other domains. Splitting the data neglects the domain commonalities and causes much less training data, making the models hard to learn. (2) Maintaining multiple models cause a tremendous amount of resource consumption and require much more human cost. It is unduly burdensome when the number of business domains is up to hundreds.  
\begin{figure}
    \centering
    \includegraphics[width=.98\columnwidth]{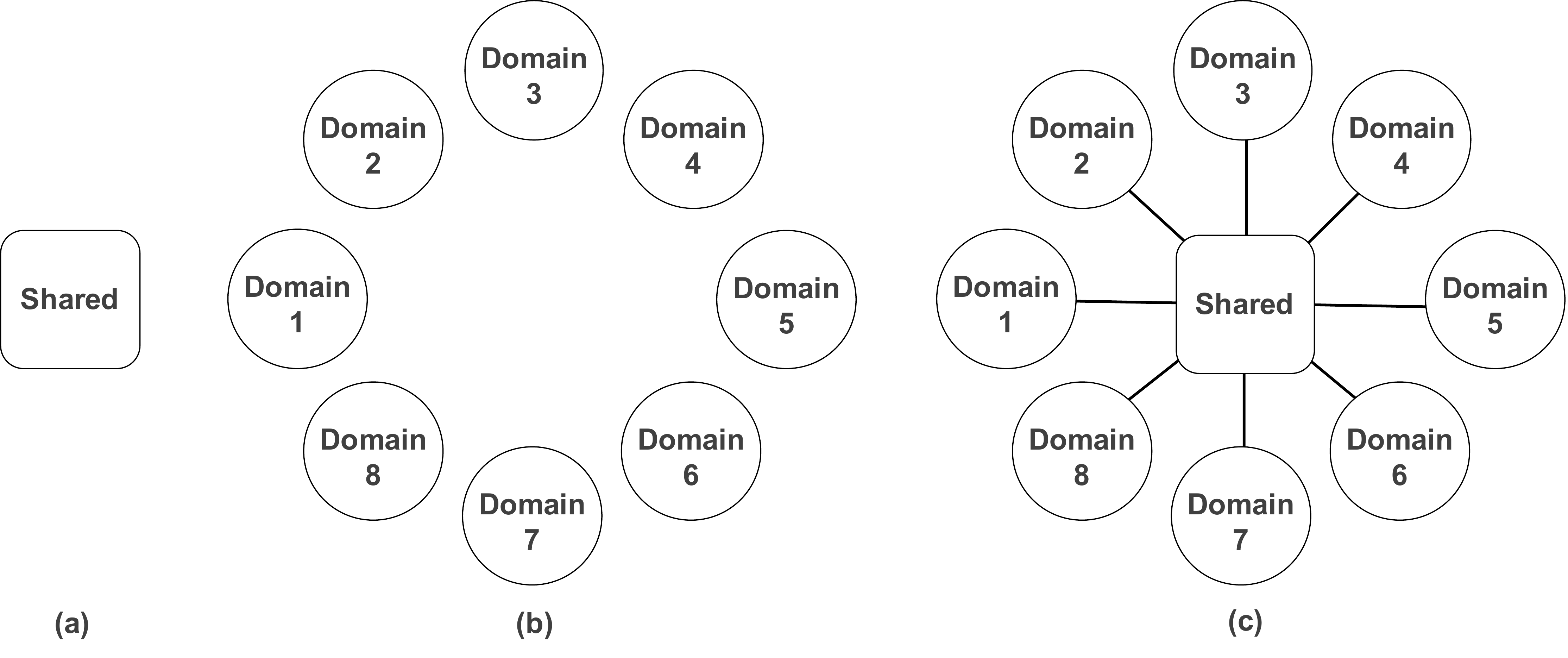}
    \caption{(a): Single shared model for all domains, square nodes indicate the shared model. (b): One model per domain where each model is learned separately. The circle node indicates the domain-specific model. (c): The proposed Star Topology Adaptive Recommender (STAR), where each domain has specific parameters and also shares a common centered model. The edges mean the combination of the center shared parameters with the domain-specific parameters.}
    \label{fig:star_topo}
\end{figure}
This paper aims to learn an effective and efficient CTR model to handle multiple domains simultaneously.
We formulate \textit{multi-domain CTR prediction} as the problem that the recommender needs to make CTR prediction for $M$ business domains $D_1, D_2, \dots, D_M$ simultaneously. The model takes input as $(\x, y, p)$, where $\x$ is the common feature used by multiple business domains like historical user behavior, user profile feature, item feature, and context feature. $y \in \{0, 1\}$ is the clicked label, and $p$ is the domain indicator that indicates which domain this sample is collected. Note that $(\x, y)$ is drawn from the domain-specific distribution $D_p$, and the distribution varies with different domains. 
Multi-domain CTR prediction aims to construct an effective and efficient model that gives accurate CTR prediction for each domain and at a trivial cost on resource consumption. To achieve this goal, the model should make full use of the domain commonalities and capture the domain distinction.

\begin{figure}
    \centering
    \includegraphics[width=\columnwidth]{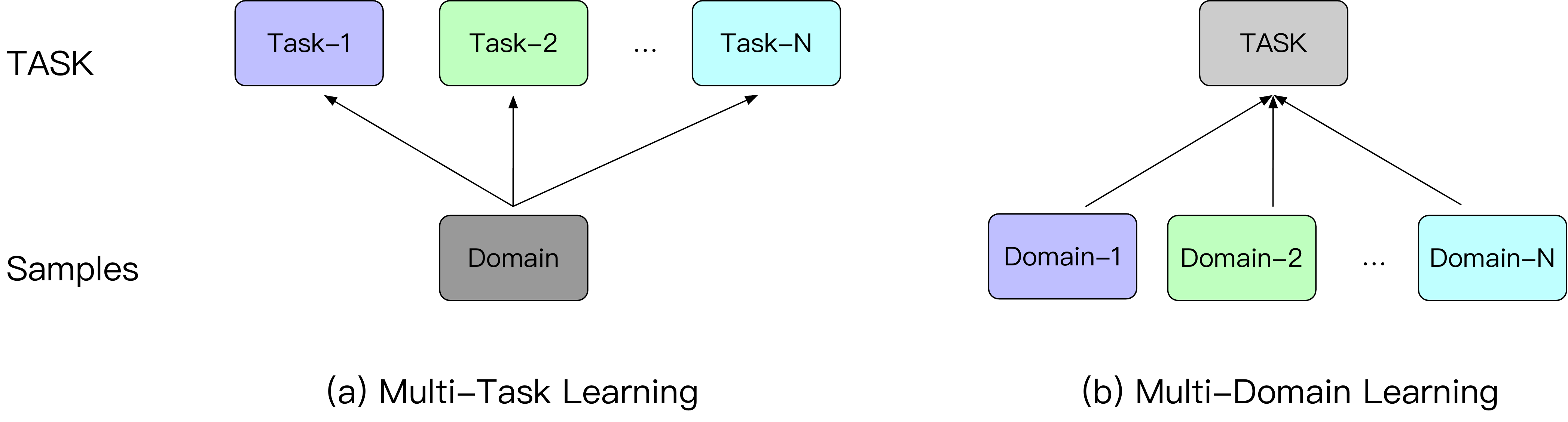}
    \caption{Comparison of multi-task learning with multi-domain learning. Most multi-task learning approaches focus on tackling different tasks within a single domain. In contrast, multi-domain learning makes predictions for multiple domains addressing the same task, e.g., CTR prediction, where the label spaces are of the same. Directly adapting multi-task approaches to multi-domain CTR prediction can not sufficiently exploit the domain relationship in the label space and neglects the distinct data distribution of different domains.}
    \label{fig:multi_task}
\end{figure}

One possible strategy to improve learning with multiple domains is multi-task learning~\cite{Ruder2017MultitaskDNN,Caruana1998Multitask,MaZYCHC2018MMOE}. As shown in Figure~\ref{fig:multi_task}, the difference between multi-domain CTR prediction and multi-task learning is that multi-domain CTR prediction solves the same task, i.e., CTR prediction, across different domains, in which the label spaces of different domains are the same and the data distribution is different. By contrast, most multi-task learning approaches~\cite{MaZYCHC2018MMOE,MisraSGH2016CrossStitch,MaZHWHZG2018ESMM,MaZCLHC2019SNR,TangLZG2020PLE} address various tasks in the same domain, where the label space might be different, e.g., jointly estimate CTR and conversion rate (CVR)~\cite{MaZHWHZG2018ESMM,WenZWLBLY2020ESM2}. Due to the heterogeneity of tasks, existing multi-task learning approaches focus on sharing information in the bottom layers but keeping separate task-specific output layers~\cite{Ruder2017MultitaskDNN}. Directly adapting multi-task approaches to multi-domain CTR prediction can not sufficiently exploit the domain relationship in the label space and neglect the distinct data distribution of different domains.

To fully exploit the domain relationship, we propose Star Topology Adaptive Recommender (\textbf{STAR}) for multi-domain CTR prediction. The proposed STAR model has the star topology, as illustrated in Figure~\ref{fig:shadow}. STAR consists of shared centered parameters and multiple sets of domain-specific parameters. The final model of each domain is obtained by combining the shared centered parameters and the domain-specific parameters. The centered parameters are used to learn general behaviors among all domains, in which the common knowledge can be learned and transferred among all domains.  The domain-specific parameters capture specific behaviors in different domains to facilitate more refined CTR prediction. The star topology facilitates effective information transformation across multiple domains to learn domain commonalities and distinctions.  
This paper implements the STAR model with the element-wise product of weights in each layer as the combination strategy.
Since embedding layers contribute most parameters in industrial recommender, the added domain-specific parameters are negligible to the total amount of parameters. Thus, using the STAR model to serve multiple domains only adds little computational and memory costs while yielding much better performance.

The main contributions of this work can be summarized as follows:
\begin{itemize}
    \item We propose Star Topology Adaptive Recommender (STAR) to tackle multi-domain CTR prediction. The star topology facilitates effective information transformation across multiple domains to learn domain commonalities while capturing domain distinction. 
    
    \item Different domains have different data distributions, this leads to inaccurate statistics when using batch normalization. We propose Partitioned Normalization (PN) that privatizes normalization for examples from different domains to address this issue. PN can lead to more accurate moments within the domain, which improves model performance. 
    
    \item  In multi-domain CTR prediction, features that depict the domain information are of importance. We propose an auxiliary network that treats the domain indicator directly as the input and learns its embeddings to depict the domain. The embeddings are then fed to the auxiliary network, which is much simpler than the original network. This makes the domain indicator influence the final prediction in a direct manner. 
    
    \item We evaluate STAR on the industrial production dataset and deploy it in the display advertising system of Alibaba in 2020. The consistent superiority validates the efficacy of STAR.  Up to now, the deployment of STAR  brings 6\% CTR and 8\% RPM lift. We believe the lessons learned in our deployment generalize to other setups and are thus of interest to researchers and industrial practitioners.
\end{itemize}

\section{Related Work}
Our work is closely related to traditional single-domain CTR prediction, where the recommender is trained on a single business domain and then serve for this business domain. Besides, our work is also related to multi-task learning and multi-domain learning. In this section, we give a brief introduction.
\subsection{Single-Domain CTR Prediction}
Inspired by the  success within deep learning, recent CTR prediction model has made the  transition from traditional shallow approaches~\cite{friedman2001greedy,rendle2010factorization,Koren08svdpp,KorenBV09MF4RS,Zhou2008ALS} to modern deep approaches~\cite{guo2017deepfm,cheng2016wide,qu2016product,zhou2018din,zhou2019dien,Pi2019MIMN}. Most deep CTR models follow the embedding and MLP paradigm. Wide \& Deep~\cite{cheng2016wide} and deepFM~\cite{guo2017deepfm} combine low-order and high-order features to improve the expression power of the model. PNN~\cite{qu2016product} introduces a product layer to capture interactive patterns between inter-field categories. In these models, the user’s history behaviors are transformed into low-dimensional vectors after the embedding and pooling. DIN~\cite{zhou2018din} employs the mechanism of attention to activate historical behaviors locally w.r.t. the given the target item, and successfully captures the diversity characteristic of user interest. DIEN~\cite{zhou2019dien} further proposes an auxiliary loss to capture latent interest from historical behaviors. Additionally,  DIEN integrates the attention mechanism with GRU to model the dynamic evolution of user interest. MIND~\cite{Li2019MIND} and DMIN~\cite{XiaoYJWHW2020DMIN} argue that a single vector might be insufficient to capture complicated pattern lying in the user and items. Capsule network and the dynamic routing mechanism are introduced in MIND to learn multiple representations to aggregate raw features. Moreover, inspired by the success of the self-attention architecture in the tasks of sequence to sequence learning~\cite{VaswaniSPUJGKP2017Transformer}, Transformer is introduced in~\cite{FengLSWSZY19DSIN} for feature aggregation. MIMN~\cite{Pi2019MIMN} proposes a memory-based architecture to aggregate features and tackle the challenge of long-term user interest modeling. SIM~\cite{PiZZWRFZG2020SIM}  extracts user interests with two cascaded search units, which achieves better ability to model lifelong sequential behavior data in both scalability and accuracy.  

\subsection{Multi-Task Learning}
Multi-task learning~\cite{Caruana1998Multitask,Ruder2017MultitaskDNN} aims to improve generalization by sharing knowledge across multiple related tasks. The shared knowledge and task-specific knowledge are explored to facilitate the learning of each task. Multi-task learning has been used successfully on multiple application domains, ranging from natural language processing~\cite{CollobertW2008MultitaskNLP}, speech recognition~\cite{DengHK2013MultitaskSpeech}, recommender system~\cite{Yuan2021OnePerson}  to computer vision~\cite{KendallGR2018MultiUncertaintyWeight}.
In early literature on MTL for linear models, \citet{ArgyriouEP2008ConvexMultitask} propose a method to learn sparse representations shared across multiple tasks. In the context of deep learning, multi-task learning is typically done with parameter sharing of hidden layers~\cite{Caruana1998Multitask,MaZHWHZG2018ESMM}. \citet{MisraSGH2016CrossStitch} propose cross-stitch units to learn unique combinations of task-specific hidden-layers for each task. \citet{MaZYCHC2018MMOE} proposes Multi-gate Mixture-of-Experts (MMoE) to model task relationships by sharing the expert sub-models across all tasks, while also having a gating network trained to optimize each task. 
\citet{KendallGR2018MultiUncertaintyWeight} propose a principled approach to multi-task deep learning which weighs multiple loss functions by considering the homoscedastic uncertainty of each task.
In multi-task learning, different tasks may conflict, necessitating a trade-off, optimize a proxy objective that minimizes a weighted linear combination of per-task losses may not be optimal. To address this issue, \citet{SenerK2018MultitaskMultiobj} explicitly cast multi-task learning as multi-objective optimization, with the overall objective of finding a Pareto optimal solution.
Note that~\cite{KendallGR2018MultiUncertaintyWeight,SenerK2018MultitaskMultiobj} are complementary to this work and could be potentially combined to achieve better performance. 

\subsection{Multi-Domain Learning}
In real-world applications, it is oftentimes that the data are collected from multiple domains~\cite{DredzeKC10CWComb,JoshiDCR2012WhenDomainMatters,LiLDZZ2020HMOE}.  Multi-domain learning enables knowledge transfer between domains to improve learning. As such, it contrasts with the domain adaptation (DA) problem~\cite{BickelBS2007DAdiscriminate,Ben-DavidBCKPV2010DATheory}, where knowledge transfer is only one way, i.e., from  the source domain to the target domain. \citet{WangJLWJ2019TransNorm} propose Transferable Normalization in place of existing normalization techniques for domain adaptation and reveals that BN~\cite{IoffeS2015BN} is the constraint
of transferability.

Multi-domain CTR prediction can be seen as a special kind of multi-domain learning problem, in which each domain corresponds to a business domain and the task is the CTR prediction. Compared with traditional multi-domain learning, our work focuses on  CTR prediction. The proposed model makes full use of the domain indicator that is directly fed as the ID feature and learning its semantic embeddings to facilitates the model learning, which is neglected by previous literature. The difference between multi-domain learning and multi-task learning is that multi-domain learning makes prediction for multiple domains addressing the same problem, e.g., CTR prediction, where the label spaces are of the same. In contrast, multi-task learning focuses on tackling different  problems~\cite{YangH2014UnifiedMultidomain}. For example, in the field of video recommendation, a multi-task learning problem can be as simultaneously predicting CTR and expected watch time of videos for a single business domain and multi-domain CTR  prediction makes CTR predictions for multiple business domains, e.g., multiple video platforms. 
\section{The Proposed Approach}
\begin{figure}
    \centering
    \includegraphics[width=\columnwidth]{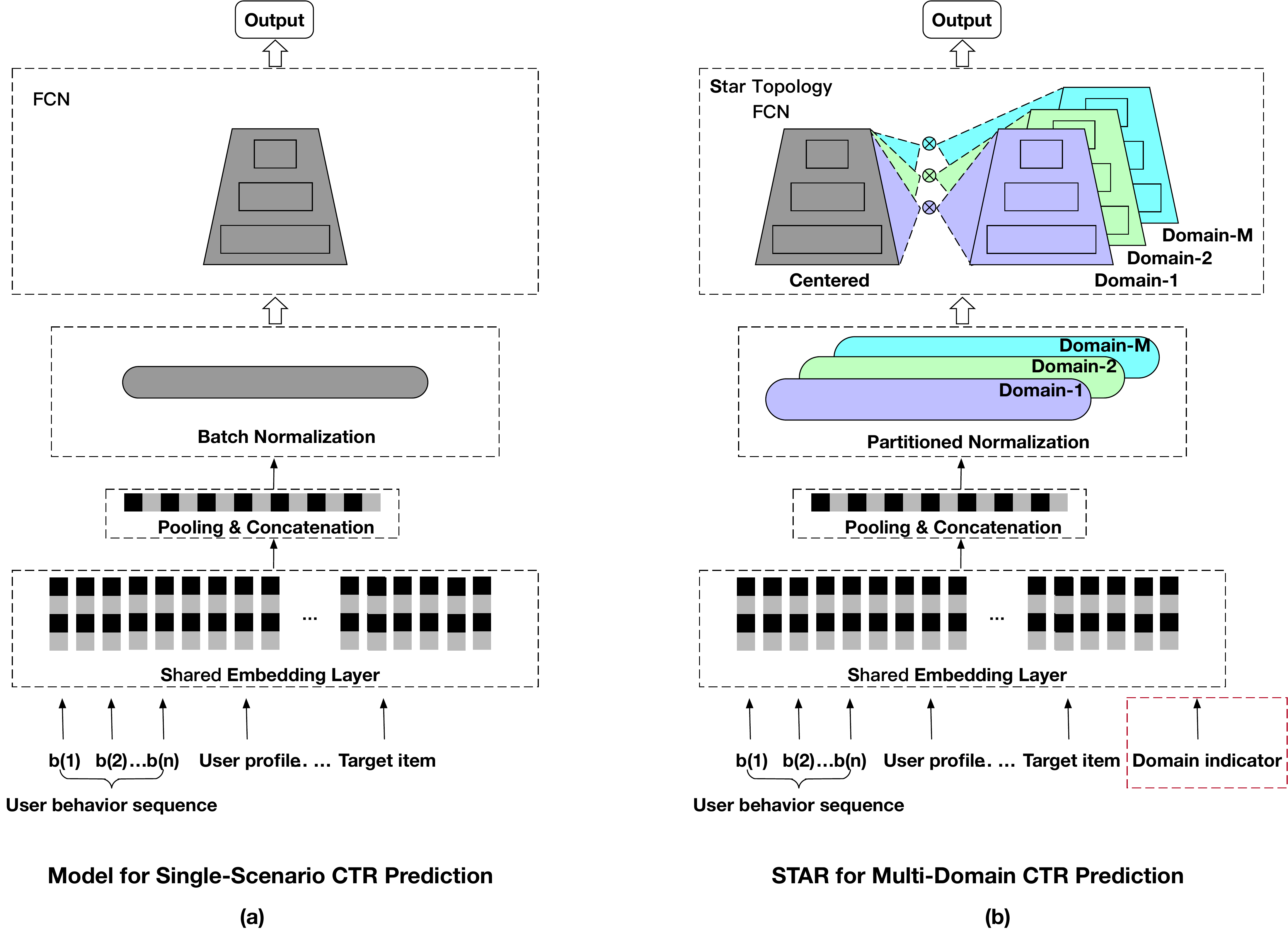}
    \caption{Comparison of model for single-domain CTR prediction and the Star Topology Adaptive Recommender (STAR) for multi-domain CTR prediction. In STAR, the partitioned normalization (PN) privatizes normalization for examples from different domains. The normalized features are then fed as input to the following star topology fully-connected neural network (star topology FCN). The star topology FCN consists of shared centered FCN and multiple domain-specific FCNs. The final combined model of each domain is obtained by the element-wise product of weights in each layer.}
    \label{fig:shadow}
\end{figure}
In this section, we first give a brief introduction about the background of multi-domain CTR prediction. Next is the architecture overview of the proposed method, star topology adaptive recommender (STAR) for multi-domain CTR prediction. 
Then we introduce STAR in detail, including the proposed star topology  network, partitioned normalization, and auxiliary network. 
\subsection{Multi-Domain CTR Prediction}
In sequential recommender systems, the model takes input as the user historical behavior, user profile feature, target item feature, and other features like context feature.
The predicted CTR $\hat{y}$ of a user $u$ clicking on an item $m$ is calculated via: 
\begin{equation*}
    \hat{y} = \mathrm{f}(E(u_1),\dots,E(u_i);E(m_1),\dots,E(m_j);E(c_1),\dots,E(c_k)),
\end{equation*}
where $\{u_1,\dots,u_i\}$ is the set of user features including user historical behavior and user profile feature. $\{m_1,\dots,m_j\}$ is the set of target item feature and $\{c_1,\dots,c_k\}$ is the set of other features. The $E(\cdot)\in \mathbb{R}^d$ means the embedding layer which maps the sparse IDs into learnable dense vectors. 

After mapping the raw features to low-dimensional embeddings, the common practice is to aggregate these embeddings to obtain fixed-length vectors. Different kinds of aggregation methods like~\cite{zhou2018din,zhou2019dien} can be employed to aggregate these embeddings to extract user interest and get the fixed-length representation. The obtained representation is then fed into the following deep neural network, e.g., a multi-layer fully-connected network, to get the final CTR prediction.

Traditional CTR models~\cite{guo2017deepfm,lian2018xdeepfm,cheng2016wide,zhou2018din,zhou2019dien} are usually trained on data from a single business domain. However, real-world recommender often has to deal with multiple business domains. 
Concretely, the recommender needs to make CTR prediction for $M$ domains $D_1, D_2, \dots, D_M$ simultaneously. The model takes input as $(\x, y, p)$, where $\x$ is the common feature used by multiple domains like user historical behavior and user profile feature, target item feature as mentioned above. $y \in \{0, 1\}$ is the clicked label and $p \in \{1, 2, \dots, M\}$ is the domain indicator that indicates which domain this sample is collected. Note that $(\x, y)$ is drawn from the domain-specific distribution $D_p$ and the distribution varies for different domains. 
The goal of multi-domain CTR prediction is to construct a single CTR model that can give accurate CTR prediction to serve all domains at low resource consumption and human cost.

\subsection{Architecture Overview}

As mentioned above, ignoring domain indicator $p$ and learning a single shared CTR model neglect the domain differences. This leads to inferior model performance. On the other hand, training separate models for each domain performs much worse since splitting the domains provides much less data for each model. Besides, it is infeasible to maintain each domain a separate model in production due to the resource consumption and human cost. 

To this end, we propose Star Topology Adaptive Recommender (STAR)  for multi-domain CTR prediction to better utilize the similarity among different domains while capturing the domain distinction. 
As shown in Figure~\ref{fig:shadow}, STAR consists of three main components: (1) the \textbf{partitioned normalization} (PN) which privatizes normalization for examples from different domains, (2) the \textbf{star topology fully-connected neural network} (star topology FCN), (3) the \textbf{auxiliary network} that treats the domain indicator directly as the input feature and learns its semantic embeddings to capture the domain distinction.

During training, a domain indicator $p$ is first sampled and then a mini-batch of $B$ instances $$(\x_1, p), (\x_2, p), \dots, (\x_B, p)$$ is sampled from this domain. STAR first embeds these input features as low-dimensional vectors by an embedding layer. In industrial recommender, the model is often trained with billions of features~\cite{Jiang2019XDL} and the parameters of embedding are usually much more than other parts of the model. This makes it difficult for different domains to learn domain-specific embeddings with limited data. For example, for models used in our daily tasks, the embeddings parameters are 10,000 times more than the parameters of fully-connected layers~\cite{Jiang2019XDL}.
Thus, in the proposed STAR model, we let all business domains share the same embedding layer, i.e., the same ID features in different domains share the same embedding. 
Sharing embedding layer across multiple domains can significantly reduce the computational and memory cost.

The embeddings are then pooled and concatenated to obtain $B$ fixed-length representations. After that, the $B$ extracted representations are processed by the proposed partitioned normalization (PN) layer that privatizes normalization statistics for different domains. 
The normalized vectors are then fed as input to the proposed star topology FCN to get the output. The star topology FCN consists of shared centered FCN and multiple domain-specific FCNs. The final model of each domain is obtained by combining the shared centered FCN and domain-specific FCN. 

In multi-domain CTR prediction, features that depict the domain information is of importance. In the STAR model, the auxiliary network treats the domain indicator as input and fed with other features depicting the domain to the auxiliary network. The output of the auxiliary network is added with the output of the star topology FCN to get the final prediction.
We make the auxiliary network much simpler than the star topology FCN to let the model capture the domain distinction in a direct and easy manner. In what follows we will describe these components in detail.

\subsection{Partitioned Normalization}
As mentioned above, the raw features are first transformed into low-dimensional embeddings and then pooled and aggregated to get the intermediate representation. Denote the intermediate representation of an instance as $\z$, to train deep networks fast and stably, a standard practice is applying normalization layer to the intermediate representation $\z$. 
Among all normalization methods, batch normalization (BN)~\cite{IoffeS2015BN} is a representative method that is proved to be crucial to the successful training of very deep neural networks~\cite{IoffeS2015BN,RadfordMC2015UnsupervisedGAN}. 
BN uses a global normalization for all examples, which accumulates  normalization moments and learns shared parameters across all samples. Concretely, the normalization of BN in training is given as  
\begin{equation}
    \z^{\prime} = \gamma\frac{\z - \mu}{\sqrt{\sigma^2+\epsilon}} + \beta,
\end{equation}
where $\z^{\prime}$ is the output, $\gamma, \beta$ are the learnable scale and bias parameters, $\mu, \sigma^2$ are mean and variances of current mini-batch.
During testing, moving averaged statistics of mean $E$ and variance $Var$ across all samples are used instead
\begin{equation}
    \z^{\prime} = \gamma\frac{\z - E}{\sqrt{Var+\epsilon}} + \beta.
\end{equation}
In other words, BN assumes all samples are i.i.d. and use the shared statistics across all training samples.

However, in multi-domain CTR prediction, samples are only assumed to be locally i.i.d. within a specific domain. Thus, data from different domains have different normalization moments. Sharing global moments and parameters of BN layers during testing will obscure domain differences and lead to degraded model performance. To capture the unique data characteristic of each domain, we propose partitioned normalization (PN) which privatizes normalization statistics and parameters for different domains. Concretely, during training, suppose the current  mini-batch is sampled from the $p$-th domain, we compute the mean and variances of the current mini-batch and normalize the feature as:
\begin{equation}
    z^{\prime} = (\gamma*\gamma_p)\frac{z - \mu}{\sqrt{\sigma^2+\epsilon}} + (\beta+\beta_p),
\end{equation}
where $\gamma, \beta$ are the global scale and bias, and $\gamma_p, \beta_p$ are the domain-specific scale and bias parameters. For each mini-batch, it receives the final scale by element-wise multiplying the shared $\gamma$ with the domain-specific $\gamma_p$, i.e., PN adaptively scales the representation according to the domain indicator. Similarly, the bias of PN is also adaptive conditioned on the domain, which is implemented by the addition of global bias $\beta$ and domain-specific bias $\beta_p$.
Note that compared with BN, PN also uses the moments of the current mini-batch during training, but PN introduces domain-specific scale and bias  $\gamma_p, \beta_p$ to capture the domain distinction. 

Besides the modification of the scale and bias, PN also let different domains to accumulate the domain-specific moving average of mean $E_p$ and variance $Var_p$. 
During testing, PN transforms instance $\z$ from the $p$-th domain as:
\begin{equation}
    \begin{aligned}
        \z^{\prime} = (\gamma*\gamma_p)\frac{\z - E_p}{\sqrt{Var_p+\epsilon}} + (\beta+\beta_p).
    \end{aligned}
    \label{eq:pn}
\end{equation}
From Equation~\ref{eq:pn}, we can see that PN uses the domain-specific mean $E_p$ and variance $Var_p$ to normalize the intermediate representation $\z$. Thus PN adaptively alters the intermediate representation conditioned on the domain indicator to capture the distinctive characteristics.


\subsection{Star Topology FCN}
\begin{figure}
    \centering
    \includegraphics[width=.8\columnwidth]{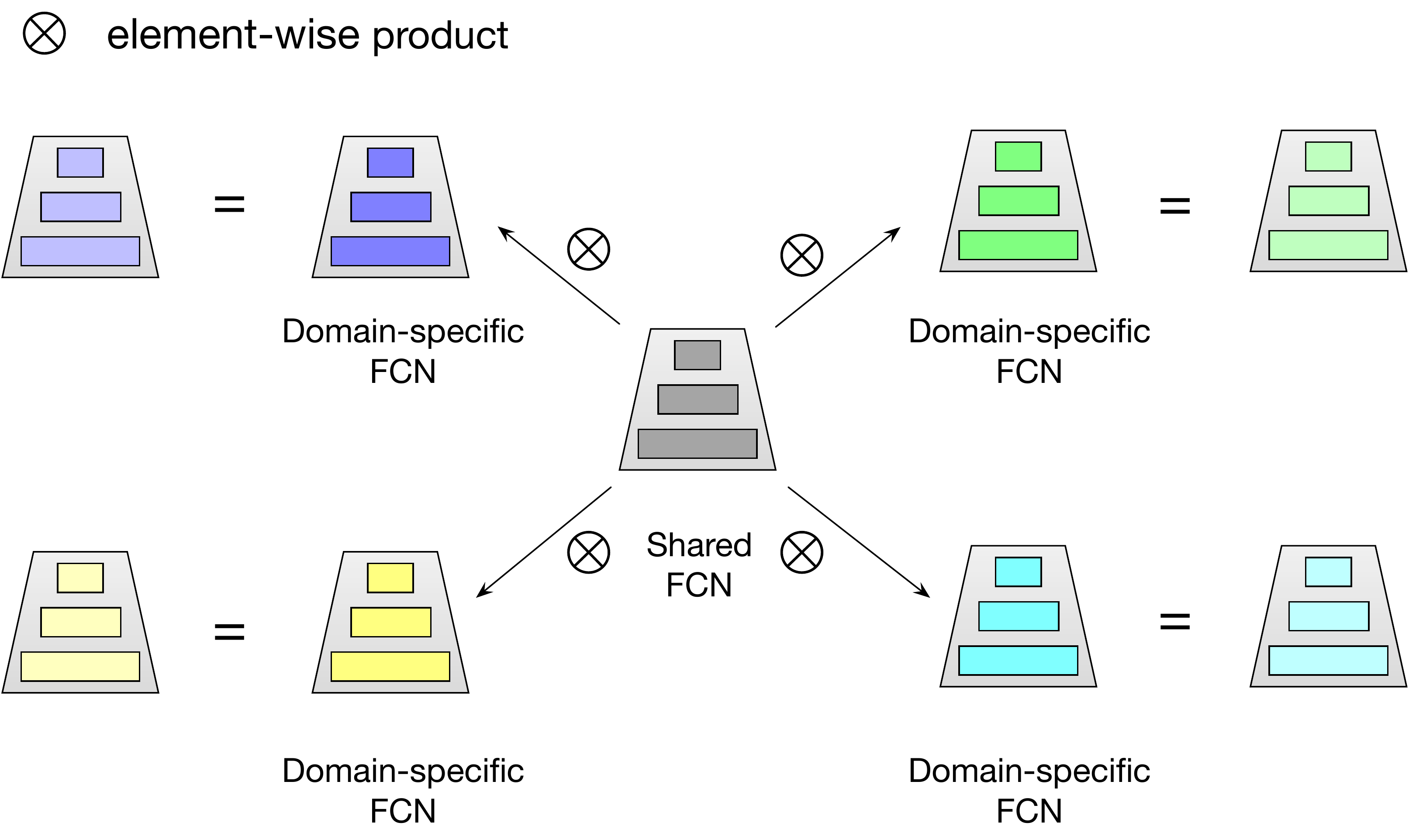}
    \caption{An illustration on how STAR generates the parameters of fully-connected network (FCN) for different domains. STAR consists of a shared centered FCN and independent FCNs per domain. For each domain, the final weights of a neural network layer are obtained by element-wise multiplying the weights of the shared FCN and the domain-specific FCN. The shared parameters are updated through the gradient of all examples, while the domain-specific parameters are only updated through examples within this domain.}
    \label{fig:star_topology}
\end{figure}
After the PN layer, the representation $\z^{\prime}$ is fed as input to the following star topology multi-layer fully-connected neural network (star topology FCN).
As depicted in Figure~\ref{fig:star_topology}, the proposed star topology FCN consists of a shared centered FCN and independent FCNs per domain, thus the total number of FCN is $M+1$. The final model of $p$-th domain is obtained by combining the shared centered FCN and domain-specific FCN, in which the centered parameters learn general behaviors among all domains, and the domain-specific parameters capture specific behaviors in different domains to facilitate more refined CTR prediction. 


Specifically, for the shared FCN, let $W$ be the weights and $b$ be the bias in a  neural network layer respectively. For the specific FCN of the $p$-th domain, let $W_p$ be the weights and $b_p$ be the bias in the corresponding layer.
Denote the input dimension as $c$ and the output dimension as $d$, i.e, $W, W_p\in \mathbb{R}^{c\times d}, b, b_p\in \mathbb{R}^d$. 
The final weights $W^{\star}_i$ and bias $b^{\star}_i$ for the $p$-th domain is obtained by:
\begin{equation}
    W^{\star}_p =  W_p \otimes W, b^{\star}_p = b_p + b, 
    \label{eq:shadow}
\end{equation}
where  $\otimes$ denotes the element-wise multiplication. Let $in_p \in \mathbb{R}^{c\times 1}$ denote the  input of this neural network layer from the $p$-th domain, the final output $out_p \in \mathbb{R}^{d} \times 1$ is given by:
\begin{equation}
    out_p = \phi((W^{\star}_p)^\top in_p + b^{\star}_p),
\end{equation}
where $\phi$ denotes the activation function of this layer.  The combination of shared parameters of domain-specific parameters is employed in all layers. By this means, STAR can modulate its parameters conditioned on the domain.

Note that we implement the combination strategy of the shared centered FCN and domain-specific FCN by element-wise product of between weights and addition of bias in each layer, other strategies can also be investigated for better performance. The shared parameters are updated through the gradient of all examples while the domain-specific parameters are only updated through examples within this domain.  This helps captures the domain differences for more refined CTR prediction while learning the domain commonality through the shared centered parameters. 
As mentioned above, most of the parameters in industrial recommenders are contributed by the embedding layer, the increased $M$ FCNs is negligible to the total amount of parameters. Thus STAR uses one model to effectively serve all business domains in a  parameter efficient and memory friendly manner.


\subsection{Auxiliary Network}
In the traditional way of CTR modeling, all features are treated equally and fed to the complicated model. In multi-domain CTR prediction, however, it may be hard for the model to automatically learn the domain distinction. We argue that a good multi-domain CTR model should have the following characteristic: (1) have informative features regarding the domain characteristic (2) make these features easily and directly influence the final CTR prediction. The intuition behind is that features that depict the information of domains are of importance since it can reduce the difficulty for the model to capture the distinction among domains.

To this end, we propose an auxiliary network to learn the domain distinction. To augment informative features regarding the domain characteristic,
we treat the domain indicator directly as the ID feature input. The domain indicator is first mapped into embedding vector and concatenated with other features. The auxiliary network then computes forward pass with respect to the concatenated features to gets the one-dimensional output. 
Denote the one-dimensional output of star topology FCN  as $s_m$ and the output of the auxiliary network as $s_a$. $s_m$ and $s_a$ are added to get the final logit. Sigmoid is then applied to get the CTR prediction:
\begin{equation}
    \textrm{Sigmoid}(s_m+s_a). 
\end{equation}
In our implementation, the auxiliary network is much simpler than the main network, which is a two-layer fully connected neural network. The simple architecture makes the domain features directly influence the final prediction.

Denote $\hat{y}^p_i$ the predicted probability for the $i$-th instance in the $p$-th domain and
$y^p_i \in \{0,1\}$ the ground truth. We minimize the cross entropy loss function  between the $\hat{y}^p_i$ and label $y^p_i$ in all domains as:
\begin{align}
    \min \sum_{p=1}^{M}\sum_{i=1}^{N_p} -y^p_i\mathrm{log}(\hat{y}^p_i)-(1-y^p_i)\mathrm{log}(1-\hat{y}^p_i).
\end{align}

\section{Experiments}
\begin{table*}[!t]
	\caption{The example percentage and average click-through rate (CTR) of each domain.}
	\label{table:sample_percentage}
	\resizebox{\textwidth}{!}{\begin{tabular}{c|c|c|c|c|c|c|c|c|c|c|c|c|c|c|c|c|c|c|c}
	\toprule
	         &1 & 2& 3& 4& 5& 6& 7& 8& 9& 10& 11& 12& 13& 14& 15& 16& 17& 18& 19\\
	\midrule
	Percentage &0.99\% & 1.61\% &3.40\% &3.85\% &2.79\% &0.56\% &4.27\% &16.76\% &10.00\% &12.16\% &0.76\% &1.31\% &3.34\% &28.76\% &1.17\% &0.46\% &1.05\% &0.91\% &5.85\%   \\
	\midrule
	CTR &2.14\% & 2.69\% &2.97\% &3.63\% &2.77\% &3.45\% &3.59\% &3.24\% &3.23\% &2.08\% &12.05\% &3.52\% &1.27\% &3.75\%  &12.03\% &4.02\% &1.63\% &4.64\% &1.42\%  \\
	\bottomrule
	\end{tabular}}
\end{table*}
We evaluate the efficacy of STAR in this section. We begin by introducing the setup including the used production dataset, compared methods and implementation details in Sec.~\ref{sec:setting}.  The results and discussion are elaborated in Sec.~\ref{sec:res}. We also perform in-depth ablation studies in Sec.~\ref{sec:ablation}. Experimental results on production environment are shown in Sec.~\ref{sec:industrial}.

\subsection{Experimental Settings}\label{sec:setting}

\textbf{Dataset.} Due to the lack of public dataset on multi-domain CTR prediction, we use Alibaba production data regarding user click behavior on 19 business domains to perform the offline evaluation.  The training data is collected from traffic logs of the online display advertising system of Alibaba. Data of one day from 19 business domains are used for training and the data of the following day is used for testing. The training dataset consists of billions of examples. Table~\ref{table:sample_percentage} shows the example percentage and average CTR (\# Click/\# Impression, i.e., ratio of positive examples) of each domain in the training set.  As shown in Table\ref{table:sample_percentage}, different domains have different domain-specific data distribution, which can be reflected from the different CTR. It can be seen that domain  with the highest CTR (domain \#15) is 12.03\% while domain  with the lowest CTR (domain \#13) is only 1.27\%. In this dataset, the majority of items are available in most of the business domains while only some of users are overlapping, e.g., domain \#1 and domain \#2 have the same set of items but only have 8.52\% overlapping users.

\textbf{Compared models.}
To verify the effectiveness of the proposed approach, we compare STAR with the following models:
\begin{itemize}
    \item \textbf{Base}. We refer to Base as the model composed of embedding layer, pooling \& concatenation layer, batch normalization, and a 7-layer fully-connected network. Specifically, the pooling \& concatenation layer is based on DIEN~\cite{zhou2019dien}, which extracts user interest after the embedding layer. We mix all samples from different domains and train the base model.
    \item \textbf{Shared Bottom}. The Shared Bottom model is a multi-task model that shares the parameters of the bottom layers. In our implementation, we let the Shared Bottom share the embedding layer. Each domain will also have a specific 7-layer fully-connected network that is not shared.
    \item \textbf{MulANN}. MulANN~\cite{SebagHSSWA2019MulANN} adds domain discriminator module to the Base model. The domain discriminator classifies which domain the examples are from. MulANN adopts a adversarial loss to let the domain discriminator indiscriminates with respect to the shift between the domains.
    \item \textbf{MMoE}. MMoE~\cite{MaZYCHC2018MMOE} implicitly models task relationships for multi-task learning, where different tasks may have different label spaces. Here we adapt MMoE for multi-domain CTR prediction, where each expert is a 7-layer fully-connected network. The number of experts is equal to the number of domains. Besides, MMoE also learns gating networks per domain that takes the input features and outputs softmax gates assembling the experts with different weights. 
    \item \textbf{Cross-Stitch}. Cross-Stitch~\cite{MisraSGH2016CrossStitch} uses linear cross-stitch units to learn an optimal combination of task-specific representations. In the cross-stitch method, each domain have a 7-layer fully-connected network and the cross-stitch units are added in each hidden layer to learn task-specific representations.
\end{itemize}
To give a fair comparison, all compared methods and the STAR model are trained with the proposed auxiliary network in Sec.~\ref{sec:res}. The ablation study about the auxiliary network is performed in Sec.~\ref{sec:ablation}.

\textbf{Implementation details.} 
All models are trained with Adam~\cite{KingmaB2014Adam}, the learning rate is set to 0.001 and the batch size is 2000. We minimize the cross-entropy loss for samples from all domains to train the model.

\textbf{Metrics.} 
Area under the ROC curve (AUC) is the common metric used to evaluate the performance of CTR prediction. An variation of user weighted AUC~\cite{zhou2018din}  measures the goodness of intra-user order by averaging AUC over users and is shown to be more relevant to online performance in recommender system. It is calculated as follows:
\begin{equation}
    \textrm{AUC} = \frac{\sum_i^n \# \textrm{impression}_i \times \textrm{AUC}_i}{\sum_i^n \# \textrm{impression}_i},
\end{equation}
where $n$ is the number of users, $\textrm{impression}_i$ and $\textrm{AUC}_i$ are the number of impressions and AUC of the $i$-th user, respectively. 
We use this weighted AUC as the evaluation metric and still refer it to as AUC for simplicity. Concretely, we use the AUC of each domain and overall AUC (mixing samples from all domains to calculate the overall AUC) as the metrics.  

\subsection{Results}\label{sec:res}
We evaluate all approaches on the Alibaba production dataset. To give a fair comparison, all compared methods and STAR model are trained with the proposed auxiliary network. As illustrated in Table~\ref{table:offline_result}, the consistent improvement validates the efficacy of STAR. 
Note that the performance of MulANN is worse than the Base model, which proves obscuring domain difference hurts the modeling of  multi-domain CTR prediction. Besides, the shared Bottom model, MMoE, Cross-Stitch and STAR all achieve better overall performance than the Base model. This demonstrates the importance of exploiting domain relationship and capturing domain distinction to enhance the prediction performance.

Although the Shared Bottom, MMoE, and Cross-Stitch achieve better overall performance than the Base model, it is notable that in some domains, the AUCs of Shared Bottom, MMoE, and Cross-Stitch are worse than the Base model, e.g., domain \# 5, \#6, and \#16. We hypothesize this is because the learning of these models conflicts in different domains. In contrast, STAR avoids this issue by its star topology, where the the domain-specific parameters are only updated through examples within this domain. 
The proposed STAR model exhibits superior performance across all domains compared with the Base model. STAR also achieves consistent improvement over the Shared Bottom, which demonstrates the importance of information sharing on top specific layers for multi-domain learning, where all domains share the same label space. STAR also outperforms MMoE and Cross-Stitch, which shows the superiority of explicitly modeling domain relationships compared with implicitly modeling domain relationships by the gate networks or cross-stitch units.


\begin{table}[t]
	\caption{Results of different approaches on offline Alibaba production dataset.}
	\label{table:offline_result}
	\resizebox{\columnwidth}{!}{\begin{tabular}{c|c|c|c|c|c|c}
	\toprule
	        & Base & Shared Bottom & MulANN  & MMoE & Cross-Stitch & STAR \\
	\midrule
	\#1
	&0.6134&0.6186& 0.6143& 0.6143& 0.6183 & \textbf{0.6306}      \\
	\#2
	&0.6321&0.6320& 0.6321&0.6355 & 0.6337&\textbf{0.6417}        \\
	\#3           &0.6281&0.6293 &0.6282
	&0.6311 & 0.6307&\textbf{0.6372}        \\
	\#4 &0.6326&0.6361& 0.6333&0.6373 & 0.6372&\textbf{0.6451}        \\
	\#5
	&0.6308&0.6292& 0.6302&
	0.6336& 0.6322&\textbf{0.6388}        \\
	\#6
	&0.6378&0.6383 &0.6336&0.6412& 0.6368&\textbf{0.6494}        \\
	\#7
	&0.6305&0.6329&0.6310 &0.6340& 0.6352&\textbf{0.6410}        \\
	\#8
	&0.6297&0.6278&0.6297 &0.6330&0.6328&\textbf{0.6411}        \\
	\#9
	&0.6264&0.6283&0.6258&0.6292& 0.6278&\textbf{0.6368}        \\
	\#10
	&0.6392&0.6434&0.6375 &0.6431& 0.6278&\textbf{0.6577} \\
	\#11
	&0.6469 &0.6529&0.6445&0.6508&0.6548& \textbf{0.6719}        \\
	\#12
	&0.6506&0.6575 &0.6498
	& 0.6518&0.6570&\textbf{0.6676}        \\
	\#13
	&0.6558&0.6612 &0.6538
	&0.6603& 0.6637&\textbf{0.6739}        \\
	\#14
	&0.6362&0.6405&0.6371
	&0.6412& 0.6411&\textbf{0.6486}        \\
	\#15
	&0.6745&0.6888 & 0.6710
	&0.6787& 0.6819&\textbf{0.7021}        \\
	\#16&0.6638&0.6627&0.6517 
	&0.6634& 0.6727&\textbf{0.6901}        \\
	\#17&0.6524&0.6658&0.6499
	&0.6519&0.6575& \textbf{0.6715}        \\
	\#18 &0.6493&0.6480&0.6375
	&0.6500 & 0.6610&\textbf{0.6754}        \\
	\#19 &0.6330&0.6375&0.6306
	&0.6374 & 0.6381& \textbf{0.6476}        \\
	\midrule
	Overall AUC & 0.6364 & 0.6398 &0.6353 
	&0.6403 & 0.6415&\textbf{0.6506}        \\
	\bottomrule
	\end{tabular}}
\end{table}

\subsection{Ablation Study}\label{sec:ablation}

To investigate the effect of each component, we conduct several ablation studies.

\subsubsection{STAR Topology FCN and PN} 

\begin{table}[t]
	\caption{Ablation study of partitioned normalization (PN) and star topology fully-connected neural networks (STAR FCN). All models are trained with the proposed auxiliary network.}
	\label{table:ablation_shadow}
	\resizebox{\columnwidth}{!}{\begin{tabular}{c|c|c|c|c}
	\toprule
	  &    Base (BN)  & Base (PN) & STAR FCN (BN) & STAR FCN (PN) \\
	\midrule
	Overall AUC & 0.6364 &0.6485 & 0.6455 &  \textbf{0.6506}    \\
	\bottomrule
	\end{tabular}}
\end{table}
We analyze the influence of different components of STAR. Concretely, the separate effects of star topology FCN  and PN are investigated. We compare (a) the Base model trained with BN, (b) Base model trained with PN, (c) STAR FCN with BN and (d) STAR model (STAR FCN + PN). The result is reported  in Table~\ref{table:ablation_shadow}. 
We observe that  using star topology FCN and PN separately can  outperform the Base model. Bring them together can further boost performance. The result validates the effect of both star topology FCN and PN.

\subsubsection{Normalization}
Normalization methods are very effective components in deep learning, which have been shown by many practices to ease optimization and enable very deep networks to converge. We analyze the effect of different normalization methods including Batch Normalization (BN)~\cite{IoffeS2015BN}, Layer Normalization (LN)~\cite{BaKH2016LN} and the proposed Partitioned Normalization (PN) on multi-domain CTR prediction. BN accumulates global statistics and learns global parameters for samples from all domains. LN is a representative instance-based normalization method, which operates along the channel dimension and avoids mixing statistics for samples from different domains. 

The result is shown in Table~\ref{table:ablation_normalization}.
Our first observation is that both LN and PN outperforms BN. This observation validates that data from different domains have distinct distribution and need specific normalization. Using global normalization obscures domain differences, which will hurt performance for multi-domain CTR prediction.
We also observe that PN outperforms LN, which validates that domain-specific normalization is better than the instance-specific normalization, since PN leads to more accurate moments within the domain.

\begin{table}[t]
	\caption{Ablation study of normalization methods for multi-domain CTR prediction. STAR FCN is trained BN, LN, and PN respectively.}
	\label{table:ablation_normalization}
	\resizebox{\columnwidth}{!}{\begin{tabular}{c|c|c|c}
	\toprule
	  &STAR FCN (BN) &    STAR FCN (LN) & STAR FCN (PN) \\
	\midrule
	Overall AUC &0.6455 & 0.6463 &  \textbf{0.6506}    \\
	\bottomrule
	\end{tabular}}
\end{table}

\subsubsection{Auxiliary network} 
\begin{figure}
    \centering
    \includegraphics[width=\columnwidth]{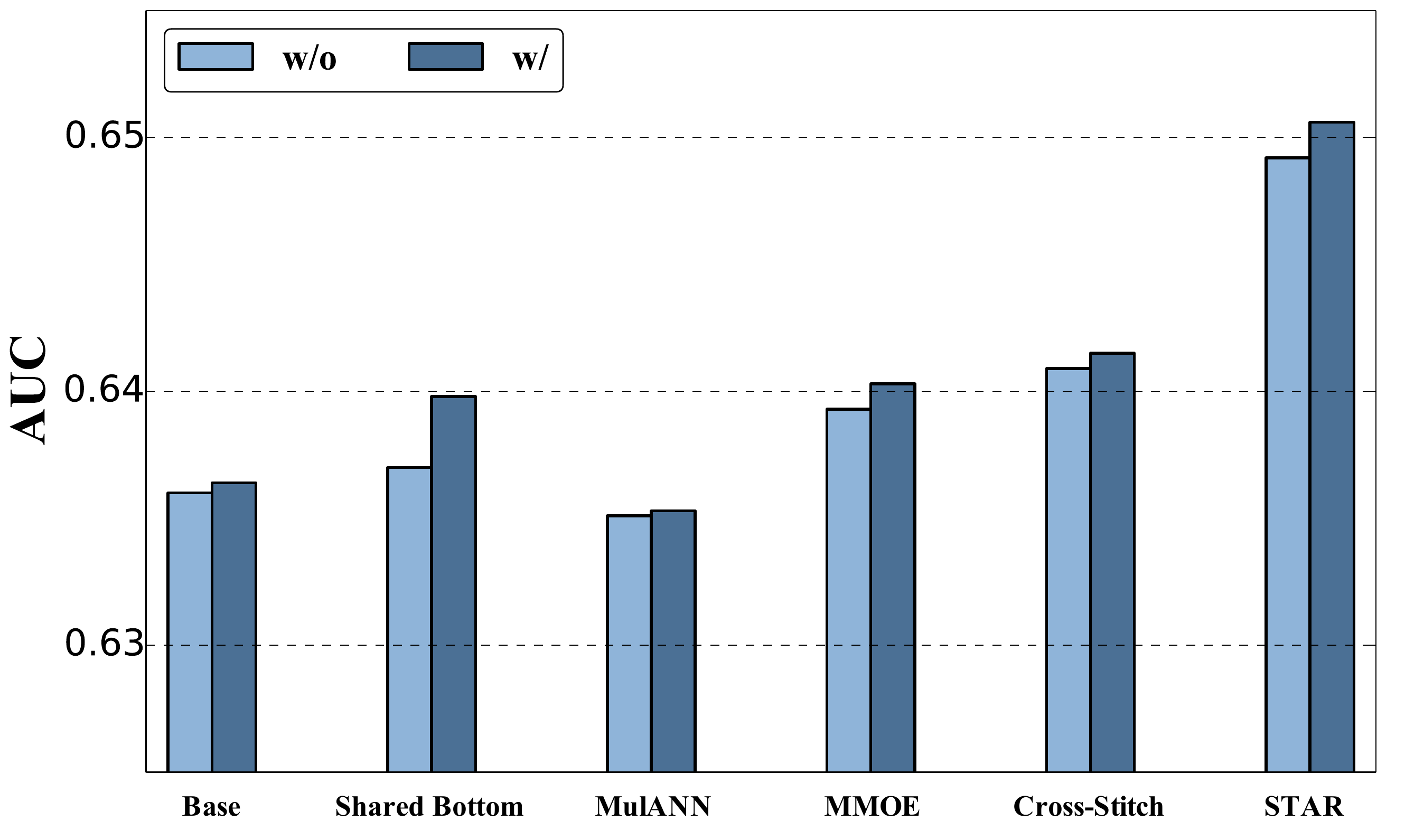}
    \caption{The performance of different methods trained with (w/) and without (w/o) the auxiliary network.}
    \label{fig:ablation_auxiliary}
\end{figure}

We conduct experiment to assess the effect of the auxiliary network for different models. All methods are trained with and without the proposed auxiliary network. The result is illustrated in Figure~\ref{fig:ablation_auxiliary}. We observe that the auxiliary network improves all methods consistently. The result validates the importance of making full utilization of domain features and using it to capture the domain distinction. We also observe the improvement of the auxiliary network for MulANN is slightly weaker than the other methods. The reason may due to the fact that the adversarial loss for obscuring domain differences contradicts with the domain feature to capture the domain differences.

\subsubsection{Ability to Capture Domain Distinction} 
\begin{figure}[!t]
    \centering
    \includegraphics[width=.8\columnwidth]{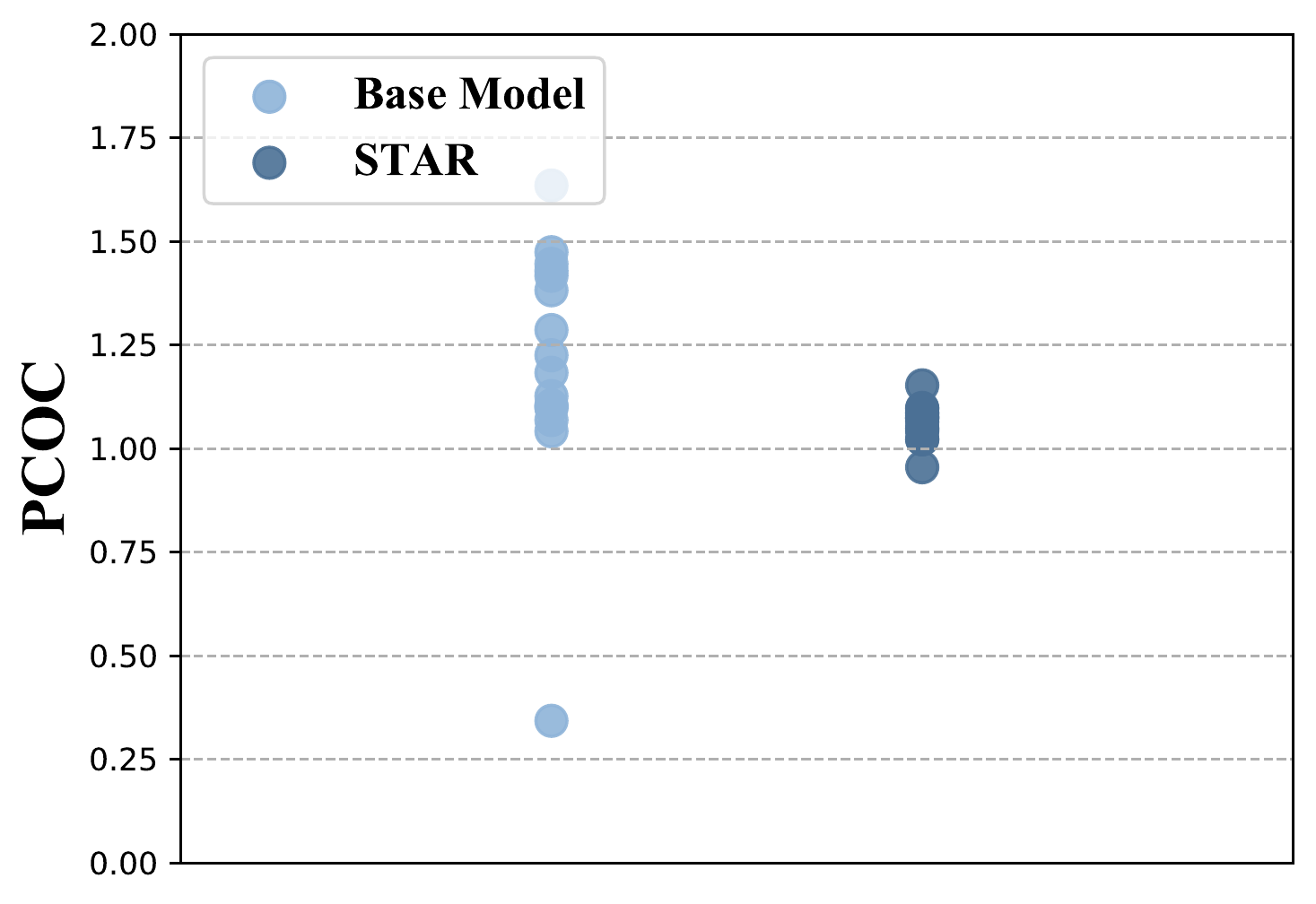}
    \caption{Predicted CTR over CTR (PCOC) of the Base model and STAR in all domains. Each circle means PCOC of a specific domain.}
    \label{fig:pcoc}
\end{figure}
Cost-per-click (CPC) is a widely used performance-dependent payment model in display advertising, where advertisers bid for clicks. In CPC, the display systems compute the effective cost per mille (eCPM) as the product of bid times its CTR. The systems allocate impressions according to the descending order of the eCPM. In CPC, the CTR model needs to be well-calibrated~\cite{GuoPSW2017OnCalibration} in order to achieve a competitive advertising system, i.e., the predicted CTR should be as close as to the actual CTR.

We show that STAR is more well-calibrated and is capable of capturing domain distinctions. We compute the predicted CTR over CTR (PCOC) in each domain. Note that the closer PCOC is to 1.0, the more accurate the CTR prediction is. For the simplicity of illustration, we show the PCOCs of the Base model and STAR in Figure~\ref{fig:pcoc}. We can see that the PCOCs of STAR in different domains are more compact and concentrated around 1.0 compared with the Base model. The result validates the ability of STAR to capture the domain distinction. 

\subsection{Production}\label{sec:industrial}
\textbf{Online serving and challenges.} 
One of the challenges in industrial recommender is that the distribution of features and CTR exhibits large shifts over time. 
To capture the dynamic change of data in real-time, it is important to use real-time examples to update the CTR models continuously to prevent them from becoming stale. 
However, for multi-domain CTR prediction, the percentage of examples of each domain changes over time. For example, some business domains have traffic spike in the morning while some business domains have traffic spike in the evening. If we train the model directly in the chronological order, the changes in data percentage over time will cause the instability of model learning. To address this issue, we redesign the data pipeline and maintain a buffer that stores a sliding window of history samples to avoid the sudden change of example percentage. Specifically, samples in the buffer are shuffled firstly and then sampled to construct a mini-batch.  
After fed to the model, this mini-batch of samples are removed from the buffer and new arriving data is added to this buffer. We empirically found this training manner is more stable than the traditional way of online updates. 

Note that during serving, the weights of FCN for each domain are pre-computed to achieve faster inferences. By this means, the computational time of STAR equals the Shared Bottom model. 
The systematical optimization makes STAR capable of serving main traffic of multiple business domains stably. Since 2020, STAR is deployed and serves more than 60 business domains on the display advertising system of Alibaba. We compute the overall improvements of all domains. Table~\ref{tab:production} shows the improvement of STAR over the previous production model, the Base model. The introduction of STAR brings +8.0\% overall CTR lift and +6.0\% overall RPM lift in our online A / B test. 
\begin{table}[t]
	\caption{CTR and RPM gains in online display advertising system of Alibaba.}
	\label{tab:production}
	\begin{tabular}{l|l|l}
	\toprule
	         & CTR & RPM \\
	\midrule
	Overall   & +8.0\%  & +6.0\% \\
	\bottomrule
	\end{tabular}
\end{table}

\section{Conclusion}
In this paper, we propose the star topology adaptive recommender to address the problem of multi-domain CTR prediction. Instead of keeping unique models for different domains or simply mixing all samples and maintaining a shared model, STAR has the star topology, which consists of shared centered parameters and domain-specific parameters. The shared parameters learn commonalities, which is updated through all examples. The domain-specific parameters capture domain distinction for more refined prediction, which is learned using examples within the specific domain. By this means, STAR can adaptively modulate its parameters conditioned on the domain for more refined prediction. The experiments demonstrate that the superiority of STAR on multi-domain CTR prediction. Since 2020, STAR is deployed in the advertising system of Alibaba, obtaining 8.0\% improvement on CTR and 6.0\% on RPM.

\bibliographystyle{ACM-Reference-Format}
\balance
\bibliography{main}


\end{document}